\documentclass[%
prb,
 amsmath,amssymb,
 reprint,%
superscriptaddress,
]{revtex4-1}

\usepackage{graphicx}
\usepackage{dcolumn}
\usepackage{bm}

\usepackage[utf8]{inputenc}
\usepackage[T1]{fontenc}
\usepackage{mathptmx}
\usepackage{xcolor}

\begin{document}

\preprint{}

\title{An atomic scale study of Si-doped AlAs by cross-sectional scanning tunneling microscopy and density functional theory}

\author{D. Tjeertes}
 \email{d.tjeertes@tue.nl}
\affiliation{Department of Applied Physics, Eindhoven University of Technology, P.O. Box 513, 5600 MB Eindhoven, The Netherlands}%

\author{A. Vela}
\affiliation{Instituto de Física, Universidade Federal do Rio de Janeiro, Caixa Postal 68528, 21941-972 Rio de Janeiro, RJ, Brazil}

\author{T. J. F. Verstijnen}%
\author{E. G. Banfi}
\author{P. J. van Veldhoven}
\affiliation{Department of Applied Physics, Eindhoven University of Technology, P.O. Box 513, 5600 MB Eindhoven, The Netherlands}%

\author{M. G. Menezes}
\author{R. B. Capaz}
\author{B. Koiller}
\affiliation{Instituto de Física, Universidade Federal do Rio de Janeiro, Caixa Postal 68528, 21941-972 Rio de Janeiro, RJ, Brazil}

\author{P. M. Koenraad}
\affiliation{Department of Applied Physics, Eindhoven University of Technology, P.O. Box 513, 5600 MB Eindhoven, The Netherlands}

\date{\today}

\begin{abstract}
Silicon (Si) donors in GaAs have been the topic of extensive studies since Si is the most common and well understood n-type dopant in III-V semiconductor devices and substrates. The indirect bandgap of AlAs compared to the direct one of GaAs leads to interesting effects when introducing Si dopants. Here we present a study of cross-sectional scanning tunneling microscopy (X-STM) and density functional theory (DFT) calculations to study Si donors in AlAs at the atomic scale. Based on their crystal symmetry and contrast strengths, we identify Si donors up to four layers below the (110) surface of AlAs. Interestingly, their short-range local density of states (LDOS) is very similar to Si atoms in the (110) surface of GaAs. Additionally we show high-resolution images of Si donors in all these layers. For empty state imaging, the experimental and simulated STM images based on DFT show excellent agreement for Si donor up to two layers below the surface. 
\end{abstract}

\maketitle

\section{\label{sec:intro}Introduction}

In the past few years, cross-sectional scanning tunneling microscopy (X-STM) has become the primary tool for performing spatial analysis of electronic wavefunctions of point defects in semiconductors \citep{Koenraad2011}. In this respect, silicon (Si) defects in III-V compounds are particularly interesting, since Si is the most common n-type dopant in these systems. 
In particular, the extensively studied III-V compound GaAs is a direct bandgap semiconductor, therefore Si donor states have strong contributions from the GaAs conduction band minimum at the $\Gamma$ point. The (110) surface of Si-doped GaAs has been well studied using X-STM \cite{Zheng1994ScanningGaAs, Domke1998Atomic-scaleSurfaces, Feenstra2002Low-temperatureSurfaces, Loth2008BandGaAs, Teichmann2008ControlledMicroscope, Garleff2011BistableArsenide, Smakman2015BistableSurface}. Signatures of Si atoms in GaAs in different charge states have been identified, and control of the charge state of surface Si donors with the STM tip and laser illumination has been shown, where the donor switches from a negative charge to a positive charge and vice versa \cite{smakman2013laser}. Si atoms below the (110) surface of GaAs are shallow donors. At negative bias they are in a neutral state (d$^0$) and at positive bias they become ionized due to tip induced band bending (TIBB), visible as rings of ionization \cite{Teichmann2008ControlledMicroscope}, leading to a positively charged donor (d$^+$). Si atoms at the (110) surface of GaAs can switch between a donor like state similar to Si atoms below the surface, and a deep DX$^-$ like state \cite{Smakman2015BistableSurface,Mooney1990DeepSemiconductors}. In the DX$^-$-like state the Si atom undergoes a bond reconfiguration and is negatively charged (d$^+$ + 2e$^-$) with an extra electron occupying the dangling bond. 

Another paradigmatic and technologically important example of a shallow donor in semiconductors is a phosphorus (P) donor in Si, where the bottom of the conduction band does not lie at center or at an edge of the Brillouin zone (BZ), but instead is located at a point in the $\Delta$ line in the BZ roughly 85\% from $\Gamma$ to $X$. The band structure then presents a 6-fold degenerate conduction band minimum, as there are six nonequivalent $\Delta$ directions.  The coupling among the 6 degenerate minima by the donor perturbation potential gives rise to interference patterns in the defect wavefunction \citep{Koiller2001}. These can be identified by STM \citep{Salfi2014, BERKOVITS1985449,Duke1996, Srivastava_1997,Baena_2016} in the (001) surface as alternating patterns, on a 4-layer cycle, depending on the impurity depth from the surface \cite{Saraiva2016DonorImages}.

Si donors in AlAs represent an interesting intermediate situation between the two aforementioned cases, as AlAs,  like Si, is an indirect gap material, but the conduction band minimum is at $X$ point, i.e. at the BZ boundary. The $X$ valley is three-fold degenerate (X$_x$, X$_y$ and X$_z$), so  the donor electron imaging on a (110) surface could potentially  reveal new symmetry features with respect to donors in direct gap materials, like GaAs, and 6-fold-degenerate indirect gap materials, like Si. Here we present such study of Si donors at various depths below a (110) surface of n-type AlAs.  

Our X-STM results are supported by electronic structure calculations. Several theoretical methods have been successfully used in the past to simulate STM images, providing a direct bridge between theory and experiment. Density functional theory (DFT) is often the preferred tool because it incorporates an {\it ab initio} description of electronic structure (including electronic correlation effects) and full relaxation of atomic coordinates, which is important when studying defects and surfaces. DFT has been used in the past to simulate STM images for reasonably deep (strongly localized) defects \cite{Capaz1995}, where relatively small supercells can be used. For shallow defects in semiconductors, like Si donors in GaAs (Bohr radius $\approx$ 100 \AA) or even P donors in Si (average Bohr radius $\approx$ 19 \AA, estimated using the geometric mean of transverse and longitudinal effective masses), the  bound electron wavefunction extends over many lattice parameters, requiring supercell sizes beyond current DFT size limits. In this case, semi-empirical methods such as tight-binding \citep{TB2007} or those based on the effective-mass theory \citep{EMT1988}  (combining DFT bulk states forcefully modulated by an empirical envelope function) \cite{Saraiva2016DonorImages}, can sometimes provide a good qualitative or semi-quantitative description of STM measurements as "cuts" of the bulk wavefunctions, without the explicit inclusion of surface relaxation or surface states.  

In this work, Si donors at different depths below the (110) surface of AlAs are identified in X-STM and their signatures are confirmed by DFT calculations. The relatively smaller AlAs average donor Bohr radius of 13 {\AA } (also estimated using the geometric mean of effective masses \cite{Vurgaftman2001BandAlloys}) allows us to obtain reliable simulated images using fully {\it ab initio} DFT calculations, which explicitly include atomically-relaxed surfaces. We obtain excellent agreement between X-STM measurements and DFT calculated images. This is, to our knowledge, the first time these techniques have been combined to study donors in III-V compounds. 

\section{Methods}\label{sec:dft}
    A Si-doped AlAs sample was grown using molecular beam epitaxy (MBE) on an n$^+$-doped (001) GaAs substrate. The sample contained (in the growth direction) a 100 nm undoped GaAs buffer, 50 nm of intrinsic AlAs, 400 nm of AlAs with $1\times10^{18}$ cm$^{-3}$ Si doping, a 50 nm intrinsic AlAs layer and a 200 nm intrinsic GaAs capping. The main purpose of the GaAs capping is to prevent the oxidation of the highly reactive AlAs layers. 
    
     For X-STM measurements, the samples are brought into the ultra-high vacuum STM chamber (typical pressure below $5\times10^{-11}$ mbar) and cleaved \emph{in situ}, revealing one of the \{110\} planes of the sample. This allows us to image a cross-section of the sample showing all the grown layers. All measurements were performed at 77 K in an Omicron LT-STM in constant current mode. STM tips were electro-chemically etched from poly-crystalline tungsten (W) wire and further tip preparation was done by sputtering with argon (Ar). STM images have their contrast adjusted individually to optimize the visibility of the features. The typical height of the atomic corrugation is 20 to 50 pm depending on the tunneling conditions. 

Our DFT calculations were performed in two steps. Initially, for smaller cells, we employ the Quantum Espresso suite (QE) \cite{Giannozzi2009QUANTUMMaterials,EnkovaaraAdvancedESPRESSO}, that uses a planewave basis to expand the electronic wavefunctions (energy cutoffs of $30$ Ry for the wavefunctions and $240$ Ry for the density). We use the local density approximation (LDA) \cite{KohnWalterMair1998, DobsonDinte1996, AndersonLangrethLundqvist1996} to describe exchange and correlation effects and ultrasoft pseudopotentials \citep{DALCORSO2014337,Prandini2018,Lejaeghereaad3000} for electron-ion interactions. The AlAs(110) surface is modeled by an 11-layer slab with a vacuum layer of 12 {\AA } to isolate the slab from its vertical periodic images. A 3$\times$4 supercell (264 atoms) of the original surface unit cell is used to accommodate the Si impurity and isolated it laterally from its periodic images. Then a Monkhorst-Pack sampling \cite{monkhorst-pack} of 3$\times$2$\times$1 is employed. All atomic coordinates are fully relaxed and then used as starting geometry for the second part of the calculations, performed with the SIESTA code \cite{siesta}, which then allows us to use larger supercells and provides an improved description of the extension and the energetics of shallow impurity states. 

In SIESTA, the wavefunctions are expanded in a pseudo-atomic DZP orbital basis, with a mesh cutoff of 300 Ry. Norm-conserving pseudopotentials were used for the ion-electron interaction and a LDA-PZ exchange-correlation functional was used for the electron-electron interaction \cite{troullier-martins, lda-pz}. The (110) surface is modeled by a slab supercell consisting of 12 atomic layers, following the labeling in Fig. \ref{fig:surface_states}. The surface supercell is now 6$\times$8 (6 zigzag chains along the [$001$] direction and each chain containing 16 atoms along the [$\bar{1} 10$] direction), resulting in a total of 1152 atoms. Additionally, a vacuum of 20 \AA \ is included. All atomic positions are allowed to relax until forces are smaller than 0.04 eV/\AA \ and the Brillouin zones were sampled by a 2 $\times$ 2 $\times$ 1 Monkhorst-Pack k-point grid \cite{monkhorst-pack}.

We simulate constant-height STM images at 2 {\AA } \ above the surface for all the systems using the Tersoff and Hamann approach \cite{tersoff1985theory}. In order to visualize the defect-level wavefunctions, the local density of states (LDOS) is integrated over narrow energy windows that include only the defect state. In QE, the windows range from $-0.18$ to $-0.11$ eV, measured from the Fermi energy, depending on the defect position. Similar windows are employed in SIESTA's calculations.

    
\section{Results and discussion}
     \subsection{\{110\} surface states of III-V semiconductors}
        \begin{figure}
            \centering
            \includegraphics[width=\columnwidth]{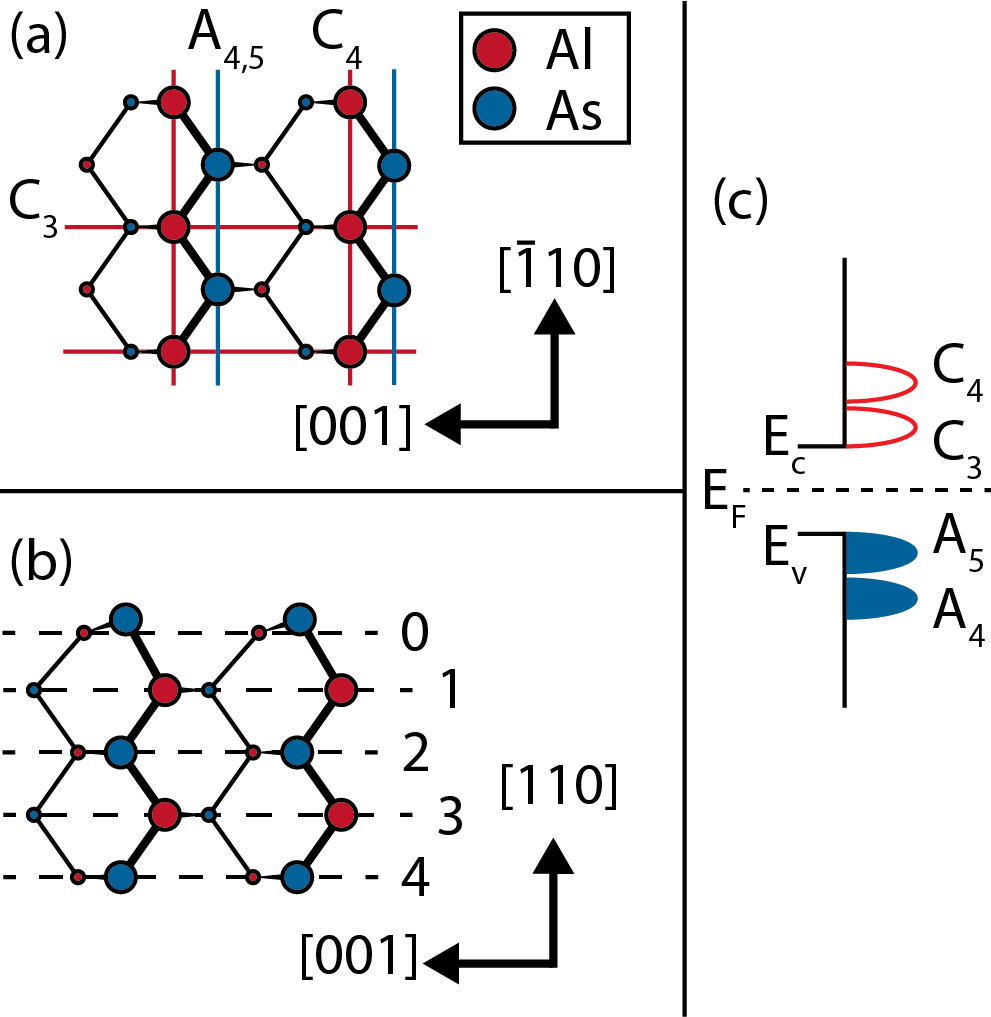}
            \caption{(a) Schematic top view of an AlAs (110) surface, with Al and As atoms colored purple and yellow respectively. The A$_4$ and A$_5$ surface states are indicated schematically with yellow lines, and the C$_3$ (horizontal) and C$_4$ (vertical) are marked with purple lines. Atoms located one layer below the surface are depicted smaller. (b) Schematic side view of the (110) AlAs surface. Layers are labelled numerically, starting with the surface as 0. In the top layer the characteristic buckling is depicted. (c) A schematic energy diagram of the surface states of an undoped III-V semiconductor with no band bending, where E$_c$, E$_F$ and E$_v$ indicate the conduction band, Fermi and valence band energy respectively.}
            \label{fig:surface_states}
        \end{figure}
        
Before presenting our results on the X-STM images of Si defects, it is interesting to review the knowledge of (110) surface states of III-V compounds, as they will set the "background" of defect state images and help us on the detailed description of the structural and electronic features of those defects. The atomic corrugation observed with STM on \{110\} surfaces of III-V semiconductors is related to a number of surface resonances, located on either side of the bandgap as schematically shown in Fig. \ref{fig:surface_states}(c) \cite{Chelikowsky1979Self-consistentGaAs, deRaad2002InterplaySurfaces}. The C$_3$ and C$_4$ states are located in the conduction band just above the band-edge, while the A$_4$ and A$_5$ states are located in the valence band, just below the band-edge. At large negative voltages, the  contrast in the STM images results from both A$_4$ and A$_5$ states. These states are constructed from particular combinations of atomic orbitals located on the anion sublattice \cite{Chelikowsky1979Self-consistentGaAs} and they both appear as rows running perpendicular to the [001] crystal axis \cite{deRaad2002InterplaySurfaces}, as shown schematically in Fig. \ref{fig:surface_states}(a). In this figure the anion (As) atoms are shown as blue disks, and the A$_4$ and A$_5$ states are represented by blue lines. Since these particular combinations of anion orbitals appear as being "connected" in the [$\bar{1}10$] direction, the image will display a series of rows along this particular direction. Shifting to smaller negative voltages moves the A$_4$ state below the tip Fermi level, leaving only the A$_5$ state which still shows, by itself, a 1D row pattern along [$\bar{1}10$] in the STM images. At even smaller negative voltages the contribution of the C$_3$ state becomes relevant. The C$_3$ state contributes to the tunneling at negative voltages due to TIBB pulling it below the Fermi level, but it only contributes significantly at small negative voltages because the contribution from the A$_5$ states is reduced substantially. The C$_3$ states are composed of p-like cation dangling bonds nearly parallel to the [001] crystal axis \cite{Chelikowsky1979Self-consistentGaAs} and therefore they contribute to STM images as 1D rows along [001], as shown schematically in Fig. \ref{fig:surface_states}(a).   Therefore, for small negative voltages, the combined contribution of both A$_5$ and C$_3$ states gives rise to corrugation along both [$\bar{1}10$] and [001] directions, i.e. 2D atomic corrugation. At small positive bias the C$_3$ state has the largest contribution to the tunnel current, resulting in a 1D corrugation running over the cation sublattice in the [001] direction. At larger positive voltages the C$_4$ states start to contribute and they are composed of combinations of cation atomic orbitals that appear in STM images as rows along the [$\bar{1}10$] direction, as indicated by the corresponding red line in Fig. \ref{fig:surface_states}(a). The STM images at these large positive voltages contains contributions from both the C$_3$ and C$_4$ states, resulting in a 2D corrugation on the cation sublattice. The exact voltages at which these surface states contribute most strongly depend on the work function of the tip, but their relative positioning with regard to voltage remains the same. 

The side view of the cleaved (110) surface is shown in Fig. \ref{fig:surface_states}(b). The layers are labeled numerically starting with the surface layer as 0 and increasing going deeper into the bulk. We adopt this labeling in all following sections. Notice that the side view is very similar to the top view, with the main difference being in the surface layer, that shows schematically the typical buckling of the (110) surface of III-V semiconductors \cite{Lubinsky1976semiconductor, Chadi1979surface, kahn1986atomic}, where the anions move slightly out of the surface compared to their ideal position, while the cations move slightly into the bulk.

\subsection{X-STM}
Fig. \ref{fig:friedel_overview}(a) shows a filled state STM image obtained on Si doped AlAs at a bias voltage of -3 V and tunnel current of 30 pA. The A$_{4,5}$ surface states can be observed as rows of bright contrast in the [$\bar{1}$10] direction. A single low frequency noise can be observed running diagonally through the image, from the bottom left to the top right. Fig. \ref{fig:friedel_overview}(b) shows the same area as (a) but the bias voltage has been changed to -1 V. The background contrast now consists of mainly C$_3$ states running in the [001] direction and the low frequency noise has disappeared. In both images a multitude of features can be observed, which will described in the following section. 
        
        \subsubsection{Identification of features based on the strength and shape of the contrast}\label{sec:ident}
            \begin{figure*}
                \centering
                \includegraphics[width=\linewidth]{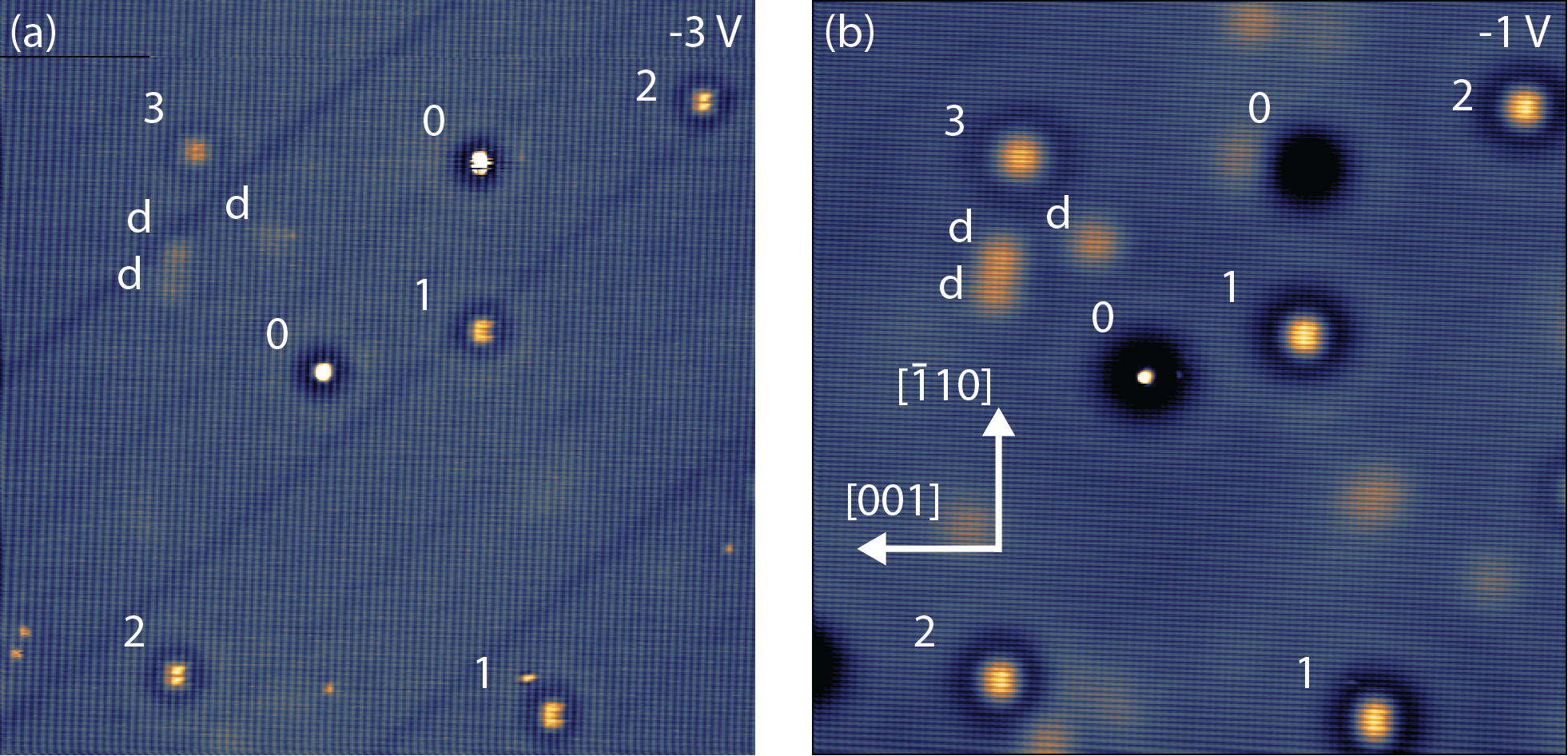}
                \caption{50 $\times$ 50 nm$^2$ filled state images obtained at two different voltages (a) -3 V, (b) -1 V containing multiple features. The numbers '1', '2' and '3' mark Si donors one, two and three layers below the surface respectively. The letters d mark Si donors located deeper below the surface and the numbers '0' mark surface defects/donors. Friedel oscillations are visible around the features and increase in amplitude when going from larger (a) to smaller (b) negative voltages. Both images were obtained with a current set point of 30 pA.}
                \label{fig:friedel_overview}
            \end{figure*}
        Si donors below the surface are marked with the numbers '1', '2' and '3', referring to their respective layer, using the labels from Fig. \ref{fig:surface_states}(b). The letter 'd' indicates Si donors deeper below the surface. The number '0' indicates either Si donors at the surface layer or other types of surface defects, which are difficult to distinguish at filled state imaging due to their very strong and localized contrast, often resulting in a virtual image of the STM tip. The donors marked by '1' and '2' show a central bright contrast with a dark ring surrounding it, which in turn is surrounded by a weaker bright contrast ring. The wavelength and amplitude of these oscillations increase at smaller negative voltages, as can be seen when comparing Fig. \ref{fig:friedel_overview}(a) (-3 V) and (b) (-1 V). These concentric rings have been attributed to charge density (Friedel) oscillations \cite{Friedel1958metallic}. 
        
        TIBB can, during filled state imaging, pull the conduction band states below the Fermi level, causing an accumulation layer at the surface. This accumulation layer can be treated as a two dimensional electron gas (2DEG), in which the Si donors act as impurities which will be screened by this 2DEG, resulting in long-range Friedel oscillations. The increase in amplitude and wavelength of the oscillations when going from large negative voltage to smaller negative voltages can be explained by the band bending. Imaging at smaller negative voltages reduces the TIBB, resulting in a local decrease of the electron density and the corresponding reduction of the distance between the Fermi energy and the bottom of the conduction band. Since the wavelength of the oscillations is proportional to the Fermi wavelength, it should decrease with decreasing Fermi energy. The reduced band bending also decreases the amount of states below the Fermi energy that contribute to the tunneling current, meaning that contribution of the electrons around the Fermi energy has increased relatively to the background, resulting in a larger amplitude of the oscillations \cite{Wielen1996DirectMicroscopy}. The closer the donor is located to the surface the stronger these oscillations will be \cite{vanderWielen1998Study}.
        
        \begin{figure}
            \centering
            \includegraphics[width=\linewidth]{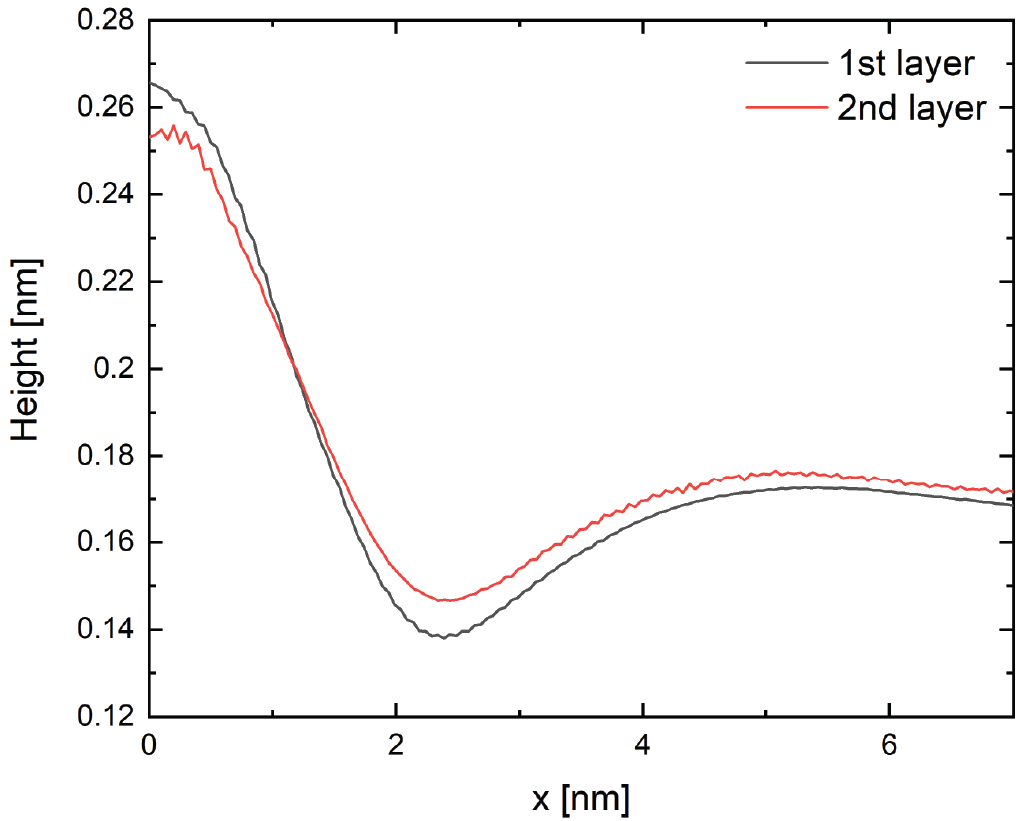}
            \caption{Radial height profiles of first and second layer donor and the Friedel oscillations surrounding them. The atomic corrugation has been filtered out. The amplitude of the height profile of the first layer Si donor is stronger than the second layer, both at the center of the feature and at the first minimum.}
            \label{fig:friedel_heigth}
        \end{figure}
        
        With this information it is also possible to determine the relative depth of the donors observed in Fig. \ref{fig:friedel_overview}(a) and (b). The Friedel oscillations are visibly stronger around the donors labeled '1', compared to those labeled '2' as shown in Fig. \ref{fig:friedel_heigth}, which shows the radial height profiles around our assigned first and second layer Si donor, where the atomic lattice has been filtered out. This leads us to the conclusion that the donors marked with '1' are indeed located closer to the surface than the donors marked with '2'. 
        
        We identified the features labeled '1' as Si donors located in the first layer (one layer below the surface), and the features labeled '2' as Si donors located in the second layer. These features are equally frequent in our images. A detailed justification for this assignment is given below. 
        
        \subsubsection{Surface layer Si donor}
            \begin{figure}
                \centering
                \includegraphics{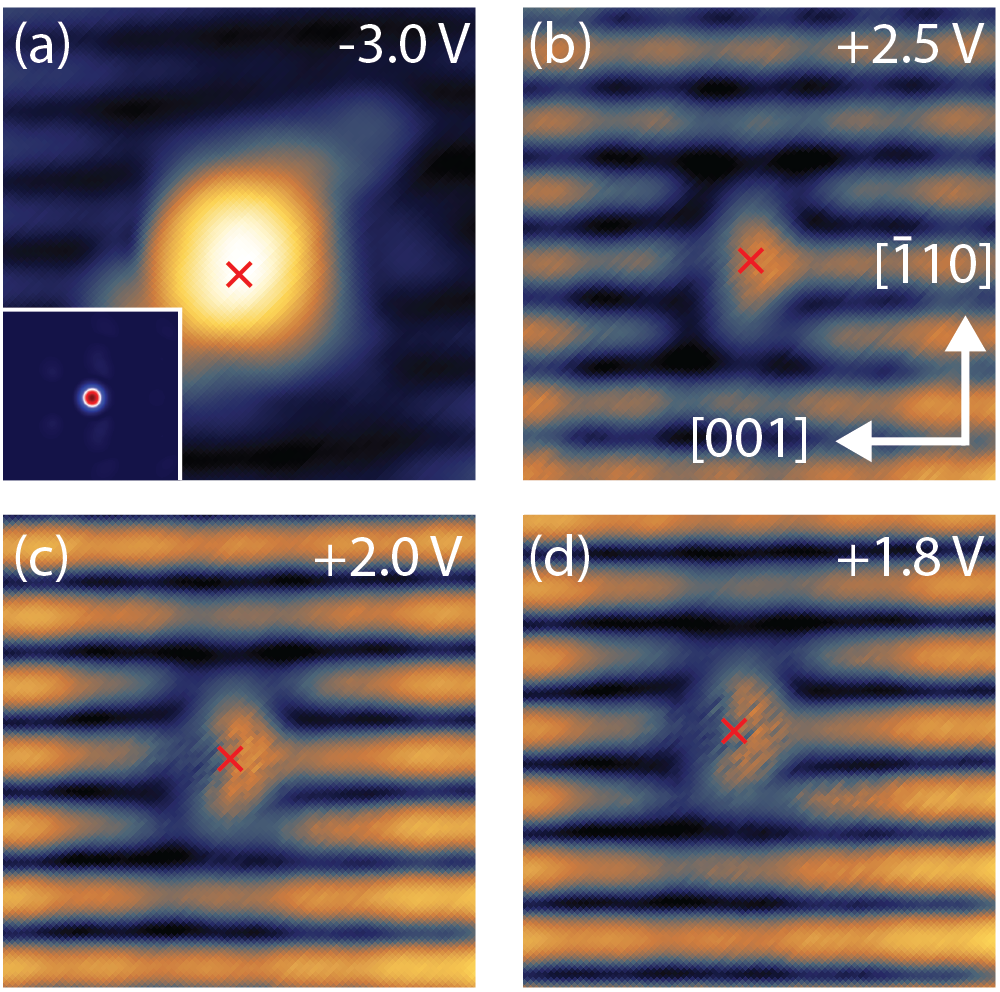}
                \caption{2.5 $\times$ 2.5 nm$^2$ images of a Si donor in the surface layer obtained at a bias voltage of -3.0 V (a), +2.5 V (b), +2.0 V (c) and +1.8 V (d). The red cross marks the location of the Si donor. All images were obtained with a set point current of 50 pA. The inset in (a) shows a 2.5 $\times$ 2.5 nm$^2$ simulated STM image of a Si$^-$ atom in the surface of AlAs.}
                \label{fig:0th_layer_voltage}
            \end{figure}
            
            Fig. \ref{fig:0th_layer_voltage}(a) shows a 2.5 $\times$ 2.5 nm$^2$ STM image of a surface Si donor obtained at a bias voltage of -3.0 V and a setpoint current of 50 pA. The contrast at the location of the feature is strong, showing an upwards tip movement of around 200 pm with respect to the corrugation. At this voltage we reason that the Si donor in the surface is negatively charged (Si$^-$), meaning that its dangling bond is filled with two electrons. Such a state has been observed for Si donors in GaAs, where it was attributed to a DX$^-$-like state \cite{Garleff2011BistableArsenide,yi2011silicon}. During filled state imaging, electrons are pulled from this dangling bond and quickly replenished from the surrounding 2DEG. Because the dangling bond is highly localized with a strong contrast, the contrast it shows is actually a convolution of the tip and the surface. This is why the feature is not perfectly symmetrical and instead its exact shape depends on the used tip (see Supplemental Material Fig. S1). The highly localized contrast is confirmed by the simulated STM image derived from DFT calculation on a Si$^-$ in the surface, which is shown as in inset in Fig. \ref{fig:0th_layer_voltage}(a). The contrast of this image ranges from blue (low), to white, to red (high).
            
            Fig. \ref{fig:0th_layer_voltage}(b), (c) and (d) show 2.5 $\times$ 2.5 nm$^2$ STM images of a surface Si donor obtained at a bias voltage of +2.5 V, +2.0 V and +1.8 V respectively. In (b) a reduction of the corrugation contrast is visible on the atoms around the feature, with a separate contrast spreading from single Al lattice site in the [$\bar{1}$10] and [1$\bar{1}$0] direction. In (c) the edges of this contrast bend in the [001] direction, creating a double lobed shaped. The two lobes lie directly on As corrugation sites. In (d) the overall strength of the contrast with respect to the corrugation decreases, and some light switching of the contrast can be observed at the location of the two lobed feature. At the positive voltages the shallow Si donor state is pulled above the Fermi level due to TIBB, and as a result is positively charged (Si$^+$).

        \subsubsection{First layer Si donor}
            \begin{figure}
                \centering
                \includegraphics[width=\linewidth]{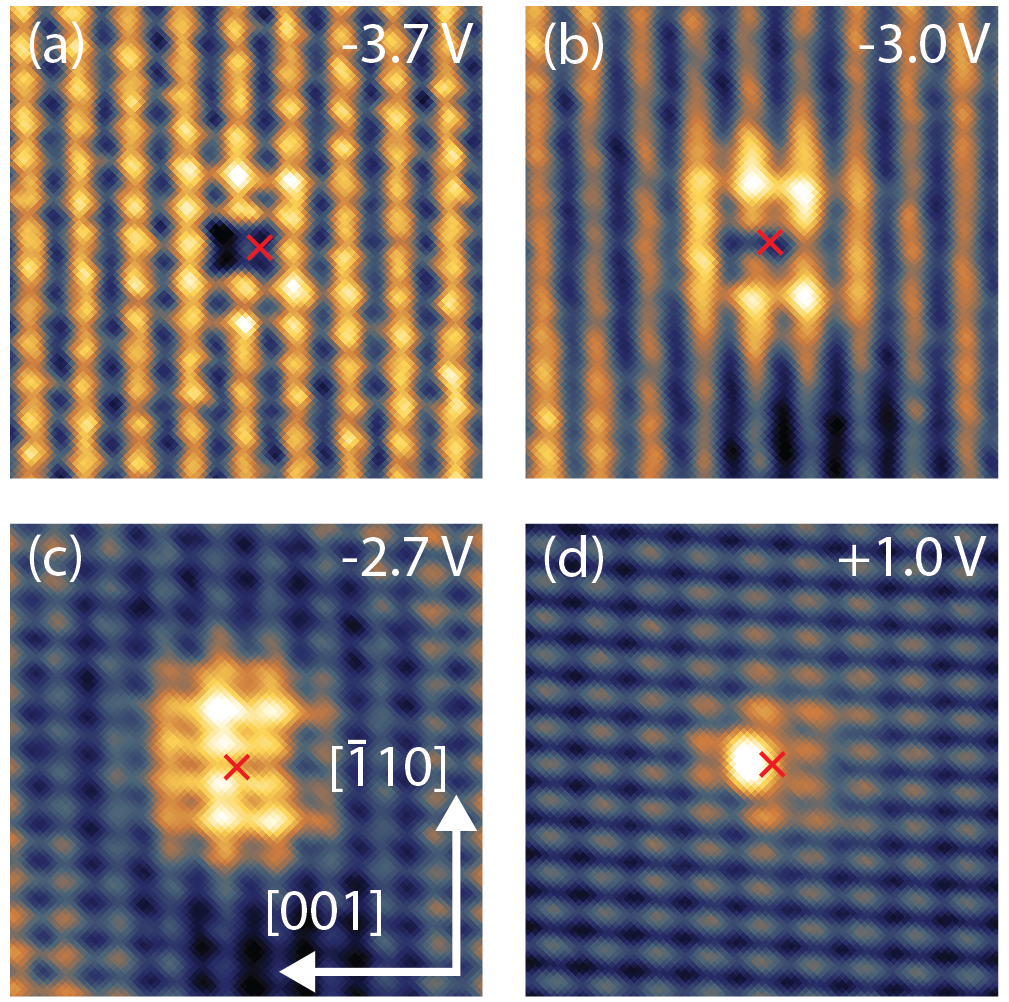}
                \caption{4.5 $\times$ 4.5 nm$^2$ images of a first layer Si donor obtained at a bias voltage of -3.7 V (a), -3.0 V (b), -2.7 V (c) and +1.0 V. The red cross marks the location of the Si donor in each of the images. All images were obtained with a set point current of 50 pA.}
                \label{fig:1st_layer_voltage}
            \end{figure}
        
Si acting as a donor in AlAs means that the Si is substitutionally incorporated at an Al site. 
At large negative voltages the As sublattice is imaged and the atomic corrugation consists of the A$_{4,5}$ states. At the location of a Si donor a depression of around 0.1 {\AA } is observed, localized on the site of a single As atom as seen in Fig. \ref{fig:1st_layer_voltage}(a). This could indicate a physical depression of the surface at this location or a reduction in the LDOS. Because of the large bias voltage we expect this to be a topographical effect. This is confirmed by our DFT calculations, which indicate a 0.2 {\AA } downward movement of the surface As atom chemically bound to the Si donor (See Supplemental Material Fig. S6). Around it, a slight increase of the contrast is observed. 
At -3.0 V the corrugation is mainly due to the A$_5$ states. The depression diminishes in strength and the bright contrast surrounding it increases in strength as shown in Fig. \ref{fig:1st_layer_voltage}(b). A ring of dark contrast now surrounds the donor, as a result of the Friedel oscillations. In Fig. \ref{fig:1st_layer_voltage}(a), due to band bending, the amount of electrons in the accumulation layer is much higher, resulting in a very short screening length, effectively removing the Friedel oscillations. 
At a bias voltage of -2.7 V the feature contains two prominent lines of bright contrast along the [001] direction, with two rows of less bright contrast in between them, as shown in Fig. \ref{fig:1st_layer_voltage}(c). The corrugation now consists of mainly A$_5$ and C$_3$ states, resulting in 2D corrugation. In between these two lines the image looks brighter than the background lattice but less bright than the two lines. Around the feature a charge density oscillation is visible. Fig. \ref{fig:1st_layer_voltage}(c) shows (at the short range) a striking similarity with a Si donor in GaAs imaged by Feenstra et al \cite{Feenstra2002Low-temperatureSurfaces}. At this voltage we reason that the shallow Si donor is neutral, d$^0$, since the downward bend bending causes the donor level to be well below the Fermi level. However, due to the tip-induced accumulation layer there will be an interplay between the extra electron from the Si donor and the 2DEG accumulation layer. This leads to a localized neutral donor like state close to the Si atom, which smoothly transitions into Friedel oscillations due to the tip induced accumulation layer at longer range. Measuring at positive bias (empty state) voltage of +1.0 V the corrugation is dominated by the C$_3$ and C$_4$ surface resonances. The Si donor now shows an enhanced contrast at 2 Al sites, but some of the anisotropic two line shape from Fig. \ref{fig:1st_layer_voltage}(c) remains. At this voltage the Si donor level is pulled above the Fermi level, causing the Si donor to be ionized (d$^+$). This is confirmed with larger scale positive bias images, where disks of ionization are visible around Si donors due to the TIBB ionizing the donor \cite{Teichmann2008ControlledMicroscope} (See Supplemental Material Fig. S2). 
        
        The observed contrast at -2.7 V is mirror symmetric, with the mirror axis parallel to the [001] direction. The mirror axis lies between two C$_3$ states, which means that the Si atom has to be located in between two of these C$_3$ states. This leads us to the conclusion that the Si atom is located either 1 or 3 layers below the surface, since in those layers the Al sites (where the Si donor will be located) lie in between the surface C$_3$ states. This is also the feature with the strongest Friedel oscillations that has this symmetry, which means that this feature can be identified as a Si donor one layer below the surface. The apparent discrepancy regarding the symmetry at -3.7 and -3.0 V can be explained by considering that the C$_3$ states are not visible at these voltages. Instead the contrast contains only A$_4$ and A$_5$ states which are centered above the As sites in both [001] and [$\bar{1}$10] direction, meaning that the mirror axis of a feature in an even layer should lie \emph{on} a corrugation row in the [001] direction. This is indeed what we observe in Fig. \ref{fig:1st_layer_voltage}(a) and (b).
        
        \subsubsection{Second layer Si donor}
        \begin{figure}
            \centering
            \includegraphics[width=\linewidth]{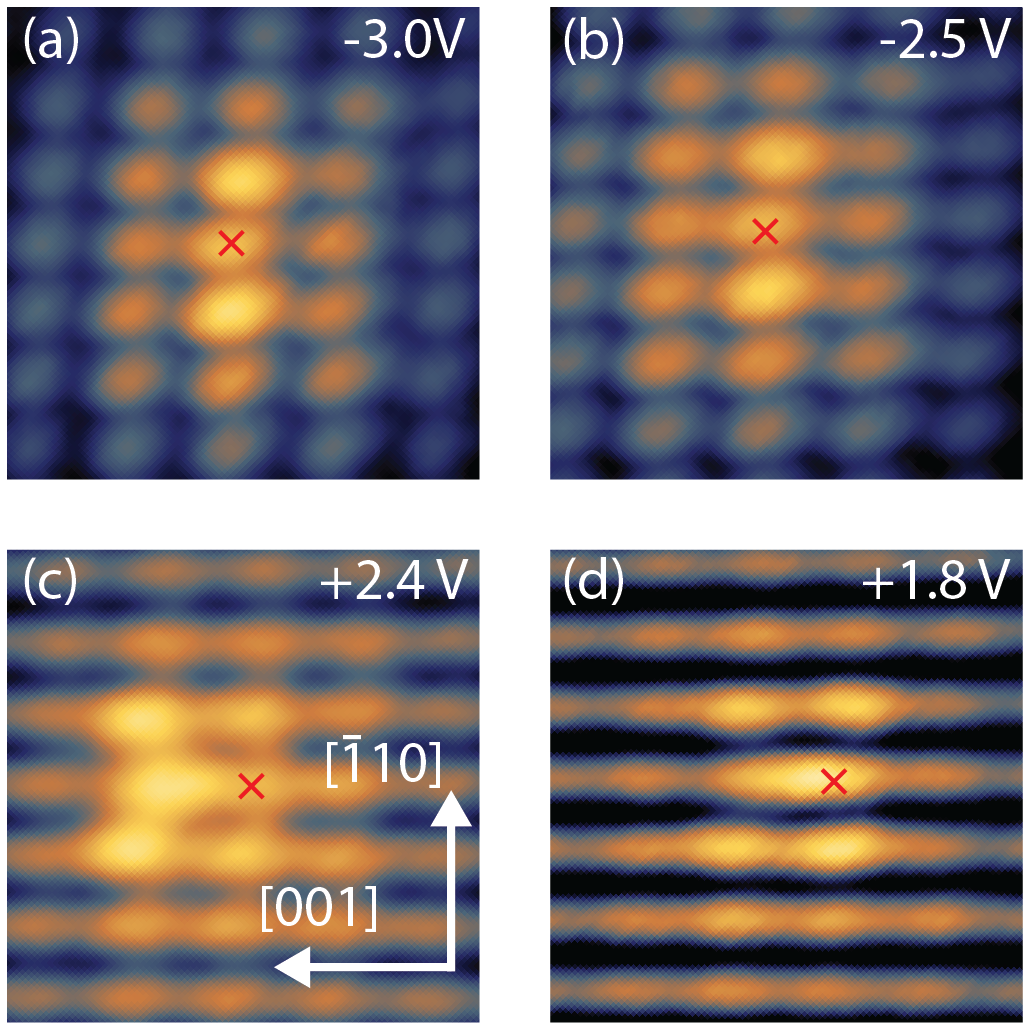}
            \caption{2.5 $\times$ 2.5 nm$^2$ images of a second layer Si donor obtained at a bias voltage of -3.0 V (a), -2.5 V (b), +2.4 V (c) and +1.8 V. The red cross marks the location of the Si donor in each of the images. All images were obtained with a set point current of 50 pA.}
            \label{fig:2nd_layer_voltage}
        \end{figure}
A Si atom in the second layer below the surface shows a two-lobed structure in filled state imaging as shown in Fig. \ref{fig:2nd_layer_voltage}(a), (b). The lobes are located along the [$\bar{1}$10] and the feature is mirror symmetric with respect to the [001] axis. In between the two lobes, the feature shows a lower brightness. When the C$_3$ surface state is contributing to the contrast for filled state imaging conditions (-3.0, -2.5 V) the location of the mirror axis is on a C$_3$ line, indicating that the Si atom is located in an even layer (0, 2, 4 layers below the surface), where the Al atoms lie directly below the C$_3$ lines. In Section \ref{sec:ident} we reasoned that, based on the Friedel oscillation amplitude, feature '2' is located deeper than feature '1', leading to the conclusion that the Si donor is indeed located 2 layers below the surface. Changing the bias voltage at filled state imaging conditions has no effect on the shape of the feature as can been seen when comparing Figs. \ref{fig:2nd_layer_voltage}(a) and (b). At this voltage there is again the interplay between the single electron on the neutral donor, d$^0$, and the surrounding 2DEG in the accumulation layer, leading to a superposition of the neutral donor image near the Si atom and the Friedel oscillations. 
        
        At a positive bias of +2.4 V, Fig. \ref{fig:2nd_layer_voltage}(c), the Si donor shows enhanced contrast on two lines in the [$\bar{1}$10] direction, with the half facing in the [001] direction having brighter contrast. The contrast surrounding these two lines is slightly brighter than the background. The brightest part of the feature is located in the [001] direction with respect to the location of the Si atom. In Fig. \ref{fig:2nd_layer_voltage}(d) the same feature is imaged with a bias voltage of +1.8 V. Now the brightest part of the feature has shifted to right above the Si atom. Again, at positive voltages the Si donor is pulled above the Fermi level and is positively charged as a result (d$^+$). Si donors in the third layer below the surface show a very similar structure to Si donors in the 1st layer, but with the feature in general being weaker and less localized in both filled and empty state images. The same holds for the Si donor that is located 4 layers below the surface. This gives additional credibility to the observed features (See Supplemental Material Fig. S3 and S4).
        
    \subsection{DFT}
        
        
        \begin{figure*}
            \centering
            \includegraphics[width = \textwidth]{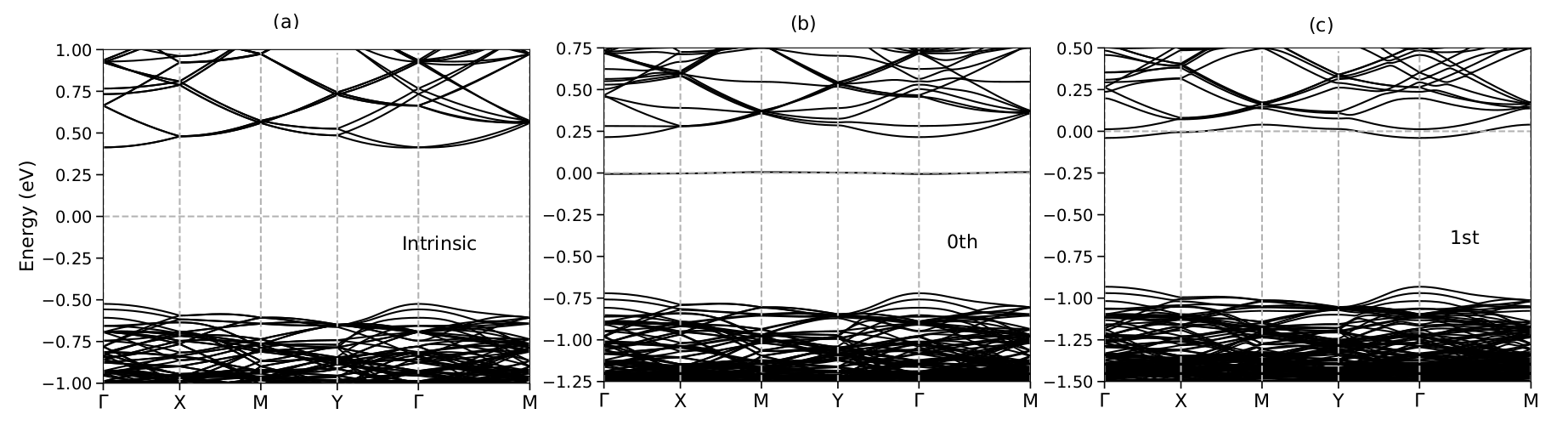}
            \caption{DFT band structure of a pristine (110) AlAs slab (a) and a Si-doped slab, with the Si donor located at the surface layer (b) and the first layer below the surface (c). The Fermi energy is set to zero in all cases. The impurity state introduced by the Si donor can be observed in the bandgap, just below the bottom of the conduction band around zero energy.}
            \label{fig:band_structure}
        \end{figure*}
        
\begin{figure}
            \centering
            \includegraphics{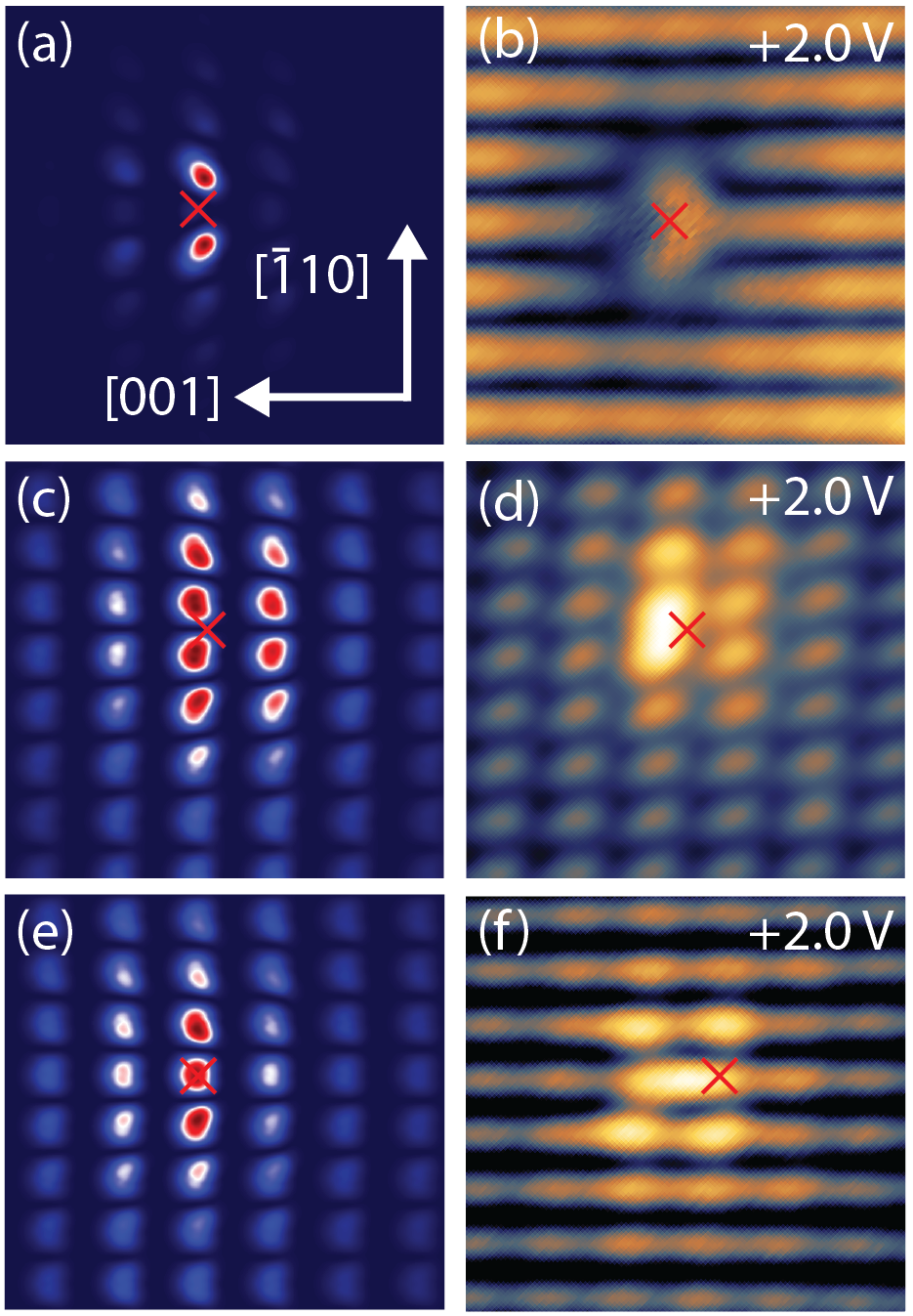}
            \caption{Simulated STM images from DFT calculations (a)(c)(e) and 2.5 $\times$ 2.5 nm$^2$ empty state STM images (b)(d)(f) of a Si donor in the surface (a)(b), one layer below (c)(d) and two layers (e)(f) below the surface. The red cross in each image marks the location of the Si donor. The color scale of the simulated STM images ranges from blue (low), to white, to red (high).}
            \label{fig:stm_dft}
        \end{figure}

In parallel with STM experiments, DFT calculations were performed on the Si-doped AlAs system.
Fig. \ref{fig:band_structure}(a) shows the calculated band structure of the (110) surface of an intrinsic AlAs slab.
In Fig. \ref{fig:band_structure}(b) the isolated impurity state introduced by the donor in the surface can be observed below the bottom of the conduction band near the Fermi level. Since these calculations are performed for the neutral donor case (d$^0$), the Fermi level is pinned by the defect state, which is half occupied. In Fig. \ref{fig:band_structure}(c) the band structure for a Si donor one layer below the surface can be observed. In this case, the impurity band can still be observed below the bottom of the conduction band around the Fermi energy, but it much shallower and the band displays a small dispersion. Consequently, the impurity state wavefunction has a greater extension in comparison with the previous case, as can be appreciated in Fig. \ref{fig:stm_dft}. Similar trends are observed when the donor is placed further below the surface, with the impurity state getting consistently closer to the bottom of the conduction band (see Supplemental Material Fig. S5).
        
DFT is also used to simulate constant-height STM images using the Tersoff-Hamann approximation \cite{tersoff1985theory}. For this purpose, we compute the LDOS for a small energy window, that includes only the defect state. 
Fig. \ref{fig:stm_dft}(a) shows the simulated STM image of a Si donor at the surface. The location of the Si donor in each image is marked with a red cross. At the selected energy range, the LDOS shows a very localized feature near the Si atom. Interestingly, we observe a two-lobed structure near the As atoms neighboring the Si atom in the surface, in striking similarity to the experimental image for positive bias, reproduced in Fig. \ref{fig:stm_dft}(b). Cross-sectional LDOS images of this defect and projected density of states (PDOS) calculations (Supplemental Material Fig. S7) show that, in addition to the two features near the As atoms, this state is also composed of a Si p-like dangling bond with stronger intensity towards the bulk (therefore harder to see in STM). Some additional and weaker LDOS can be observed in neighboring corrugation row in the [001] direction. In Fig. \ref{fig:stm_dft}(b) an empty state STM image of a Si donor in the surface obtained with a bias voltage of +2.0 V is displayed. 
Additionally the experimental image shows the C$_3$ surface states, running along the [001] axis (the simulated STM image does not contain the C$_3$ states because it is taken at a smaller energy window, as we mentioned, including only impurity state). The simulated STM image with an overlay of the atomic lattice can be found in Fig. S8 of the Supplemental Material. 
For this particular case of the surface donor, we have also performed calculations for a negatively charged impurity. We find a metastable state in which the impurity lies closer to the surface (see Fig. S12 of the Supplemental Material). The total energy of this configuration is about $0.3$ eV higher than the most stable configuration, which is identical to the one found for the neutral donor (see Fig. S6). The LDOS profile corresponding to the impurity state found in this metastable configuration, displayed in the inset of Fig. \ref{fig:0th_layer_voltage}(a), displays a strong localization on top of the Si atom, as discussed in section III.B.2. Such a metastable configuration is not found in the neutral case.

Fig. \ref{fig:stm_dft}(c) shows a simulated STM image of a Si donor one layer below the surface. The Si donor shows an enhanced LDOS at multiple rows of surface Al atoms in the [$\bar{1}$10] direction, with the stronger features being concentrated on three of these rows. The middle row shows the strongest LDOS at four Al sites, of which the middle two show the highest brightness. The neighboring corrugation row in the [00$\bar{1}$] direction shows the second highest LDOS, with again four sites appearing brighter, of which the middle two the most. The whole feature is mirror symmetric with respect to the [001] axis, with the mirror plane located in between two surface Al atoms. This defect state feature can be rationalized as a product between a delocalized Al-like conduction band surface state and an envelope function that imposes localization near the Si donor. The apparent size of this envelope function is consistent with the average donor bulk Bohr radius of $\approx$ 13 {\AA } (Supplemental Material Fig. S9) obtained from a hydrogenic model within effective mass theory. In Fig. \ref{fig:stm_dft}(d) an empty state STM image of a Si donor one layer below the surface is displayed (the same image as in Fig. \ref{fig:2nd_layer_voltage}(d)). Comparing this image with the simulated STM image obtained from DFT calculations in \ref{fig:stm_dft}(c), a very similar structure is observed. Both show enhanced LDOS at three surface Al rows in the [$\bar{1}$10] direction, with the middle one having the largest LDOS. The asymmetry on the LDOS brightness of the neighboring Al rows is also observed, with four Al sites showing enhancement in the row in the [001] direction and only two sites showing strong enhancement in the [00$\bar{1}$] direction. This provides further confirmation that the Si donor imaged in Figure \ref{fig:stm_dft}(d) is located one layer below the surface and shows that the DFT calculations provide a successful method to simulate STM images. 

The Si donor located two layers below the surface (Fig. \ref{fig:stm_dft}(e)) follows the same trend (enveloped Al surface states), showing an enhancement of the LDOS at three Al surface rows in the [$\bar{1}$10] direction. Again the middle one shows the strongest brightness, localized on three Al atoms. The same goes for its two neighboring rows, with the neighbor in the [001] direction appearing slightly brighter than the one in the [00$\bar{1}$] direction. The whole feature is mirror symmetric with respect to the [001] axis, with the mirror plane going through the central Al row. The comparison between Figs. \ref{fig:stm_dft}(e) and (f) (experiment) shows again an excellent match. Both features are mirror symmetric with respect to the [001] axis, with the mirror plane located on the central Al row, which matches the idea that the Si donor is located directly below a surface Al site in an even layer. In both simulated and experimental images the middle row is the brightest. The contrast of the Si atom three layers below the surface is shown in Supplemental Information Fig. S13. Overall there is a strong agreement between the simulated and experimental STM images for Si donors in the top four layers of the AlAs (110) surface. In the Supplemental Material, we provide extra theoretical LDOS plots showing cross-sectional features of the defect states contributing to the STM images. In these plots, the interplay between defect and surface states is elucidated.

\section{Conclusions}
    We observed Si donors in multiple layers of the AlAs (110) surface. We identified Si donors up to four layers below the surface based on their contrast. Around all Si donors Friedel oscillations are observed at small negative voltages, caused by an accumulation layer at the surface. Interestingly, the short-range character of the Si defect LDOS in AlAs is not very different from Si in GaAs.
In addition, neither the oscillatory behavior due to valley interference nor the strongly layer dependent character of the LDOS (features observed for P atoms in Si) are seen. Using simulated STM images based on DFT calculations we were able to match experimental empty state images of Si donors to the simulated STM images. The identification of the layers in which the Si atoms are located made based on the experimental results matches with the simulated STM images. Overall there is a strong agreement between the simulated and experimental STM images for donors in the top three layers of the AlAs (110) surface. This provides a good basis for further research combining STM measurements and DFT calculations on donor-like systems. 
\begin{acknowledgments}
We acknowledge financial support from Brazilian funding agencies CNPq, FAPERJ, INCT-Nanomateriais de Carbono and INCT-Informação Quântica. 
\end{acknowledgments}

\end{document}


\title{Supplemental material: \\ An atomic scale study of Si-doped AlAs by cross-sectional scanning tunneling microscopy and density functional theory}

\author{D. Tjeertes}
 \email{d.tjeertes@tue.nl}
\affiliation{Department of Applied Physics, Eindhoven University of Technology, P.O. Box 513, 5600 MB Eindhoven, The Netherlands}%

\author{A. Vela}
\affiliation{Instituto de Física, Universidade Federal do Rio de Janeiro, Caixa Postal 68528, 21941-972 Rio de Janeiro, RJ, Brazil}

\author{T. F. Verstijnen}%
\author{E. G. Banfi}
\author{P. J. van Veldhoven}
\affiliation{Department of Applied Physics, Eindhoven University of Technology, P.O. Box 513, 5600 MB Eindhoven, The Netherlands}%

\author{M. G. Menezes}
\author{R. B. Capaz}
\author{B. Koiller}
\affiliation{Instituto de Física, Universidade Federal do Rio de Janeiro, Caixa Postal 68528, 21941-972 Rio de Janeiro, RJ, Brazil}

\author{P. M. Koenraad}
\affiliation{Department of Applied Physics, Eindhoven University of Technology, P.O. Box 513, 5600 MB Eindhoven, The Netherlands}

\maketitle 

\begin{figure*}
    \centering
    \includegraphics{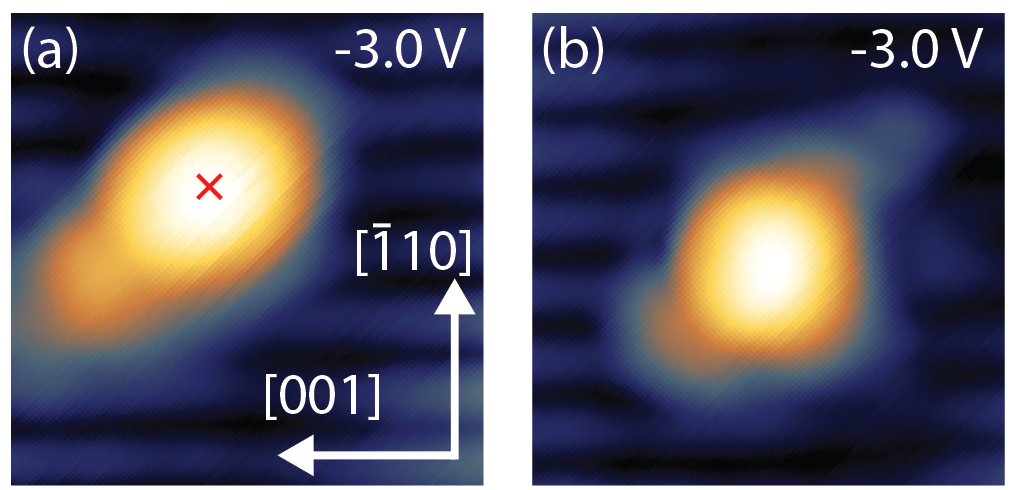}
    \caption{(a) 2.5 $\times$ 2.5 nm$^2$ filled state STM image of a surface layer Si donor. (b) Filled state STM image of the same surface layer Si donor as (a) but with an altered tip shape. Both images were obtained with a bias voltage of -3 V and a set-point current of 50 pA.}
\end{figure*}

\begin{figure*}
    \centering
    \includegraphics{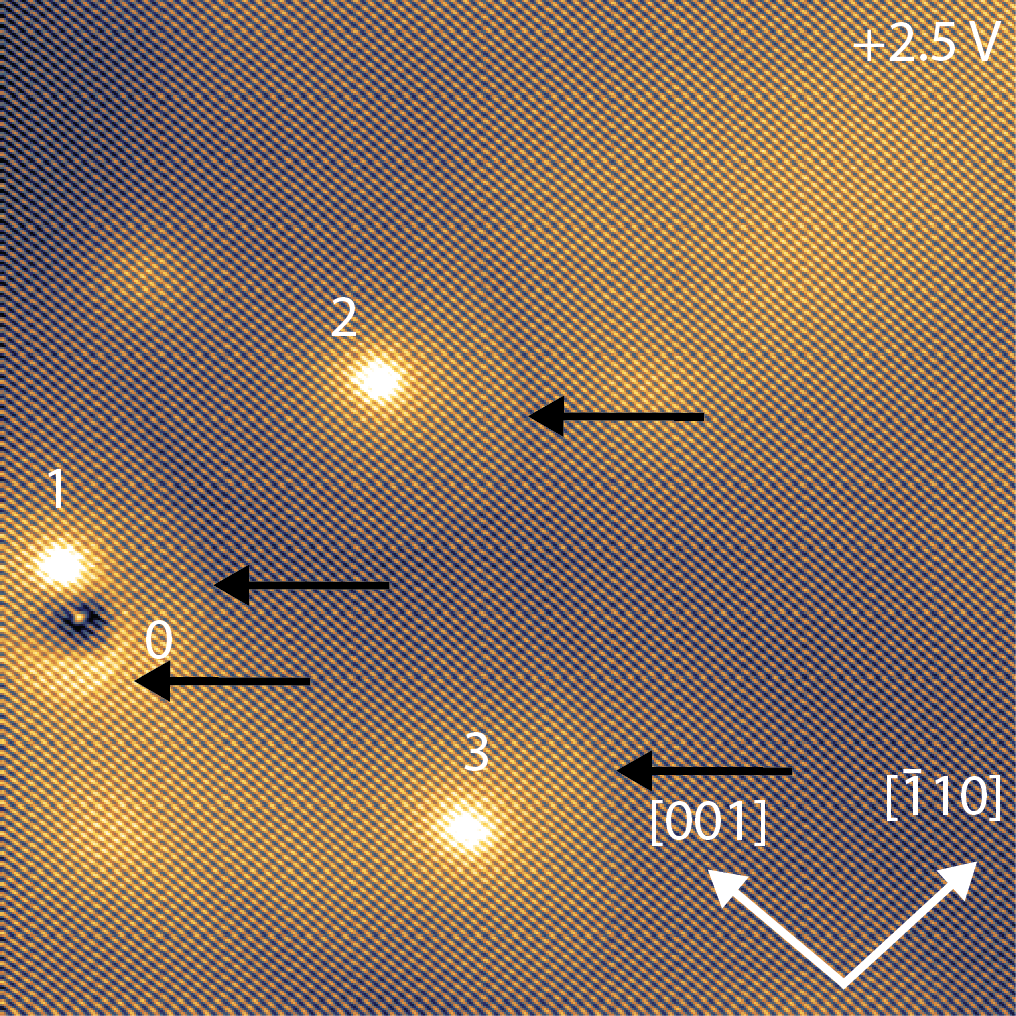}
    \caption{15 $\times$ 15 nm$^2$ filled state STM image of multiple Si donors in AlAs. Si donors are marked with the numbers 0 to 3 indicating in which layer below the surface they are located (with the surface defined as 0). Around each donor a disc of bright contrast can be observed, of which the edge is marked with a black arrow for each of the donors.}
\end{figure*}

\begin{figure*}
    \centering
    \includegraphics{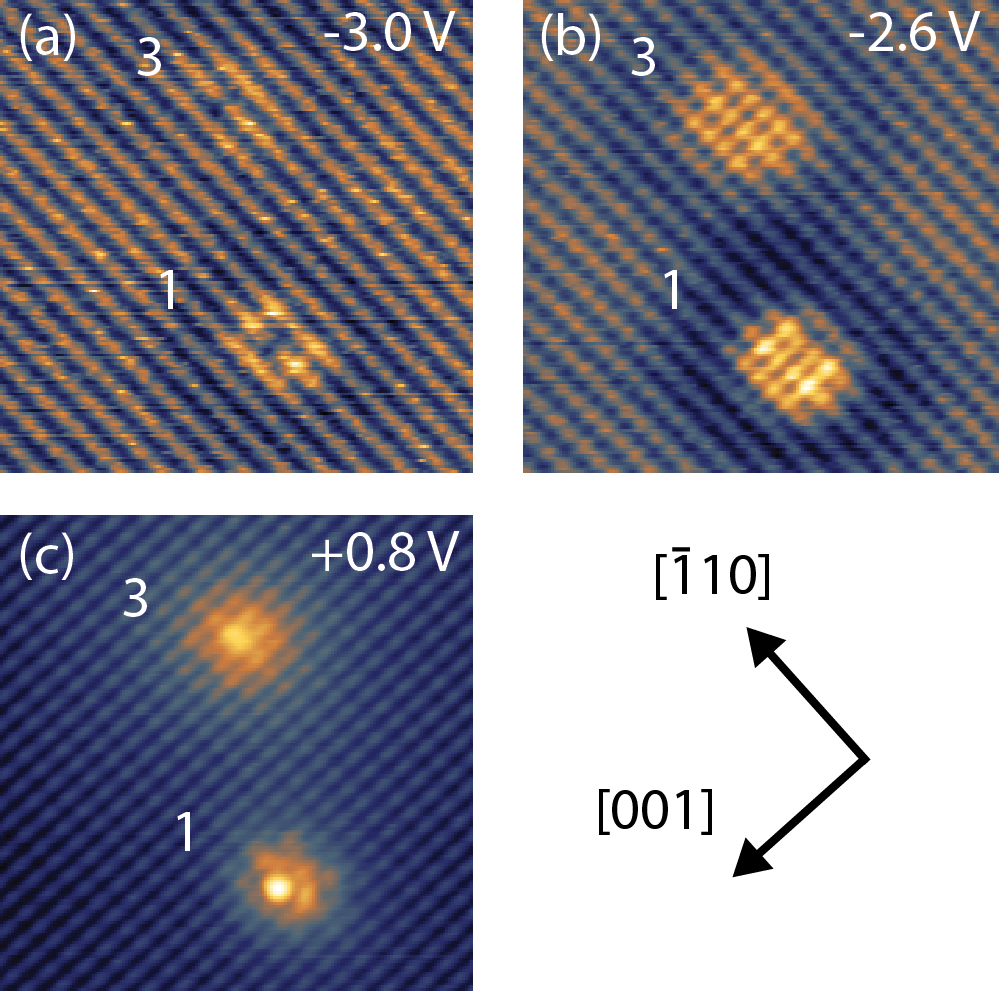}
    \caption{11.5 $\times$ 11.5 nm$^2$ STM images of an area containing a Si donor one and three layers below the surface. The Si donors are marked with 1 and 3 respectively. The images were obtained with a bias voltage of (a) -3 V, (b) -2.6 and (c) +0.8 V. All images were obtained with a set-point current of 50 pA.}
\end{figure*}

\begin{figure*}
    \centering
    \includegraphics{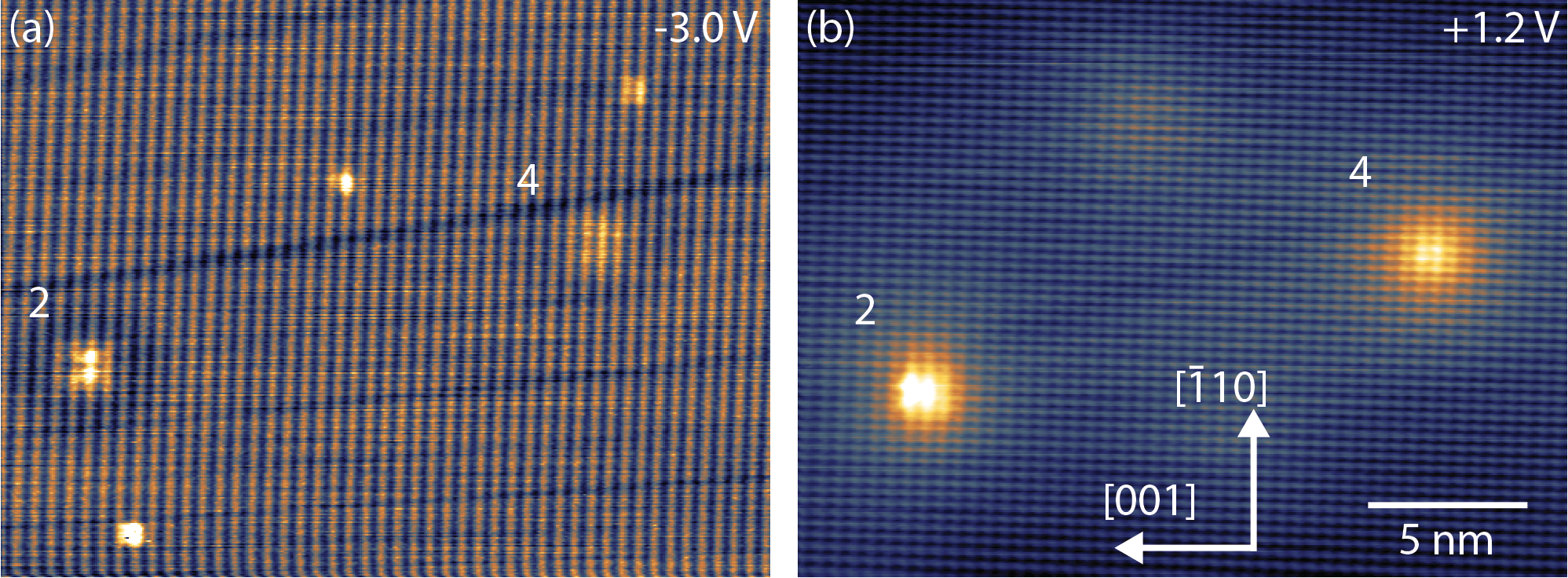}
    \caption{STM images of an area containing a Si donor two and three layers below the surface. The Si donors are marked with 2 and 4 respectively. The images were obtained at a bias voltage of (a) -3 V and (b) +1.2 V. All images were obtained with a set-point current of 30 pA.}
\end{figure*}

\begin{figure*}
    \centering
    \includegraphics[width=\textwidth]{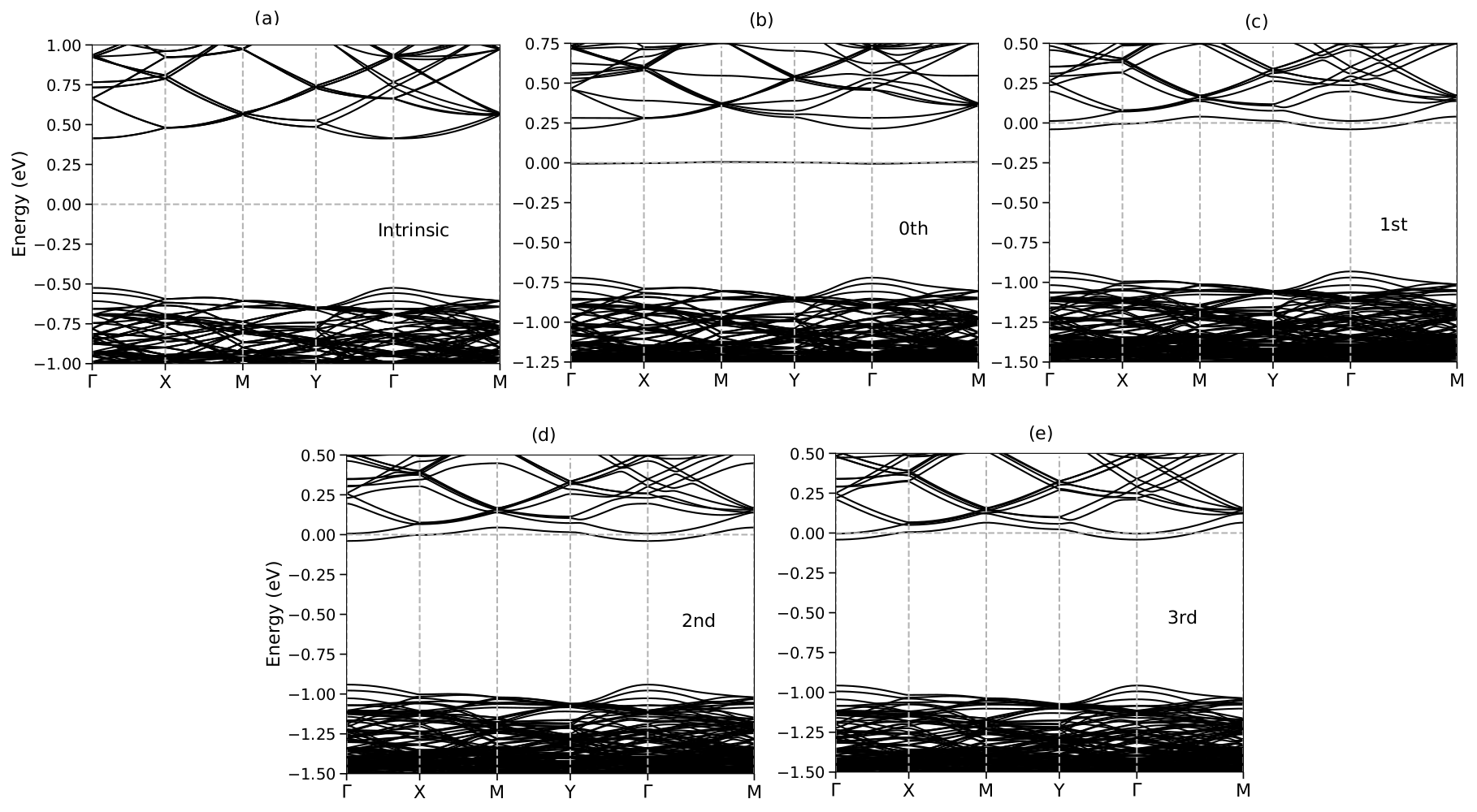}
    \caption{DFT bandstructures for a clean AlAs slab (a), and a AlAs slab with a Si donor in the surface (b) and 1 (c), 2 (d) and 3 (e) layers below the surface. This figure is an extension of Fig. 7 from the main text.}
\end{figure*}

\begin{figure*}
    \centering
    \includegraphics[width=\textwidth]{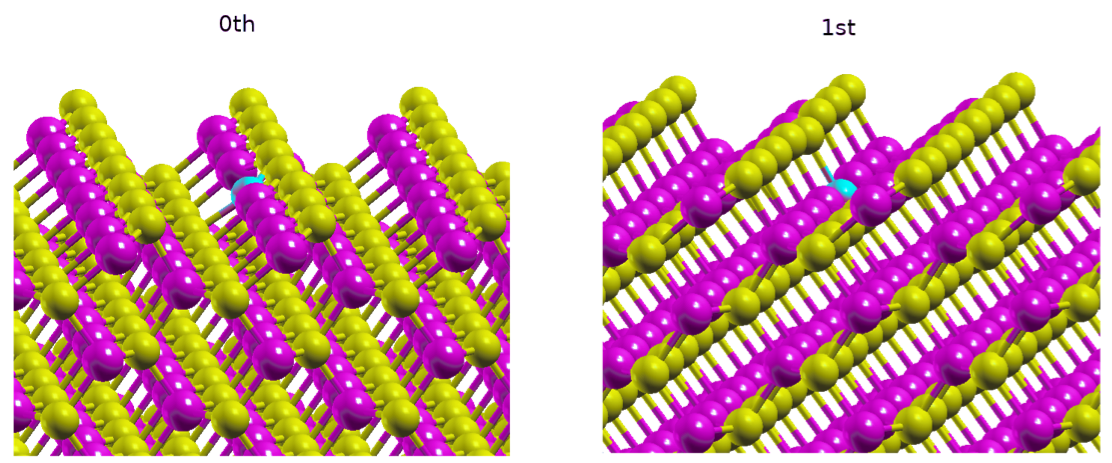}
    \caption{Part of the slab supercells for the (110) surfaces with the defect at the surface layer (left) and one layer below the surface (right) after geometry optimization. The Si atom is represented by a cyan ball and the Al and As atoms are represented by purple and yellow balls, respectively.}
    \label{fig:slab_geometry}
\end{figure*}

\begin{figure*}
    \centering
    \includegraphics[width=0.8\textwidth]{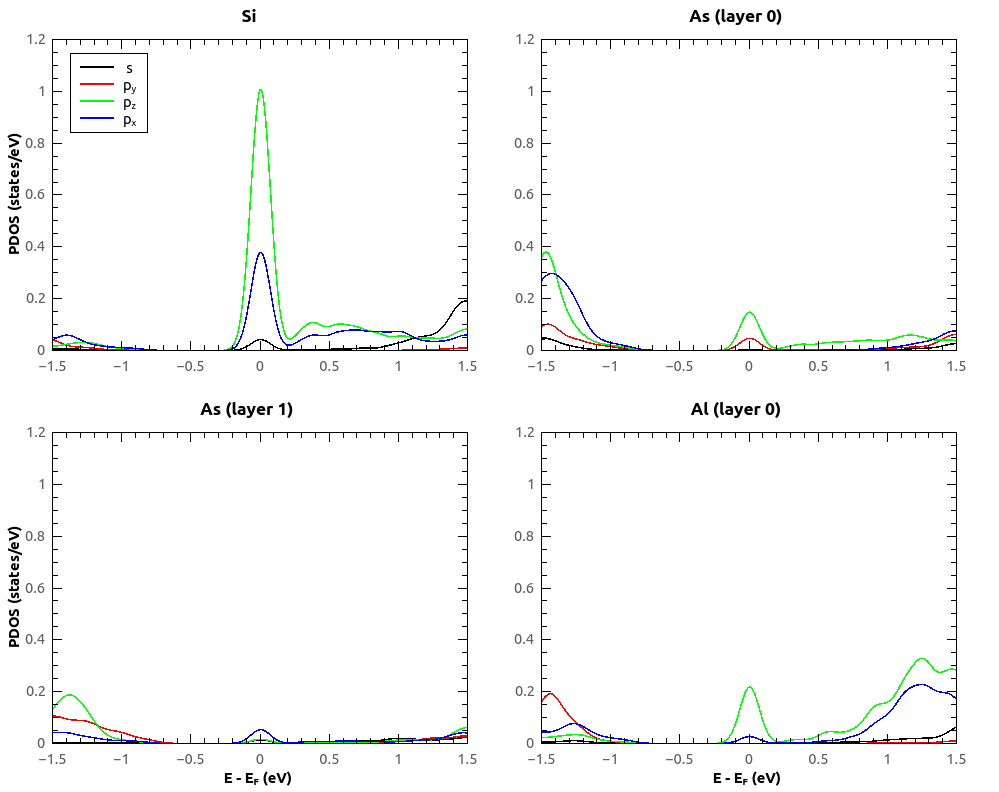}
    \caption{Projected density of states (PDOS) for the impurity atom and neighboring atoms when the impurity is placed at the surface layer (layer 0). These calculations show that the impurity level is composed of $p$ orbitals from the impurity itself and from neighboring atoms. In particular, the contributions from the As atoms can be seen in the STM images.}
\end{figure*}

\begin{figure*}
    \centering
    \includegraphics[width=\textwidth]{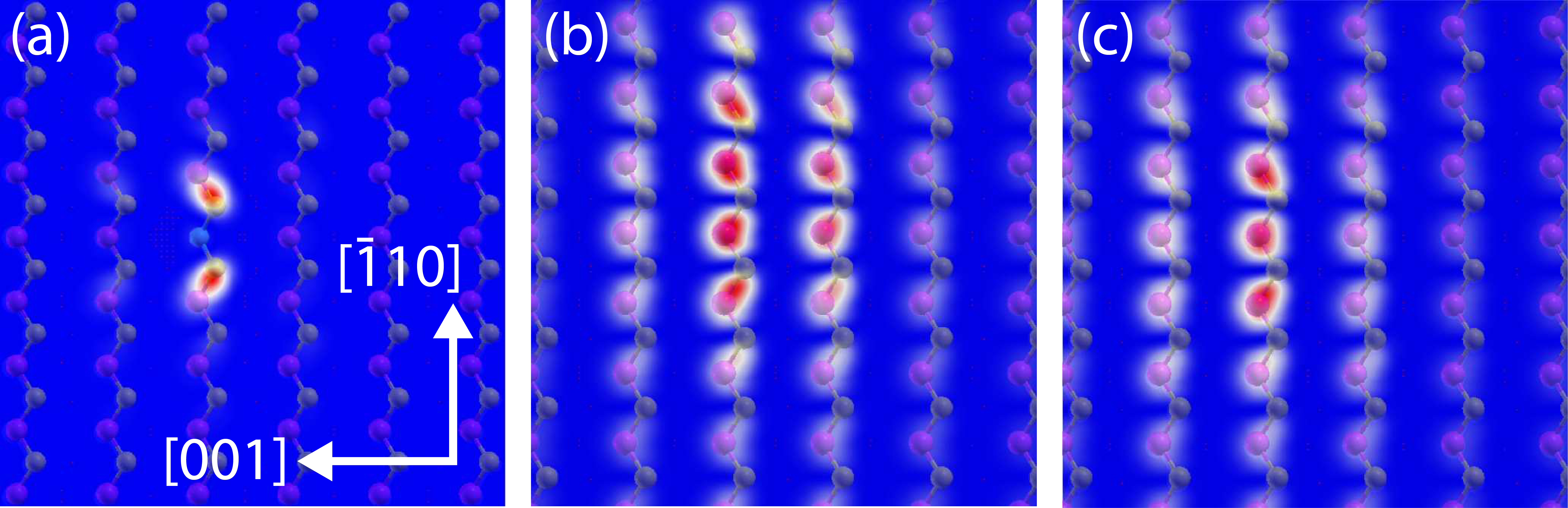}
    \caption{Simulated STM images of a Si atom in layer 0 (a), 1 (b) and 2 (c) of the (110) AlAs surface, with an overlay of the atomic lattice.}
\end{figure*}

\begin{figure*}
    \centering
    \includegraphics[width=\textwidth]{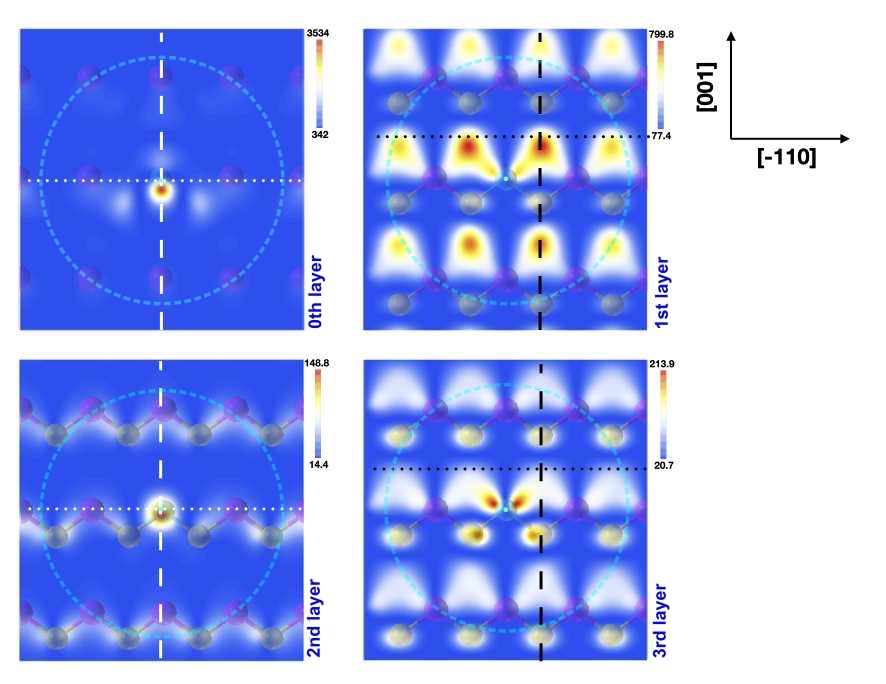}
    \caption{STM simulated images at 2$\AA$ from the surface of AlAs(110) slab for Si impurities at different depths. Cyan point shows the location of the Si impurity. The dotted and dashed white lines indicate the planes used for cross-sectional cuts in Figs. \ref{ED_001_Si_Slab} and \ref{ED_-110_Si_Slab}. The cyan circle has a 13  $\AA$ radius, the same as Bohr radius, with center in the impurity. The units are electrons per \AA$^3 \times 10^{-6}$.
}
\end{figure*}

\begin{figure*}
    \centering
    \includegraphics[width=\textwidth]{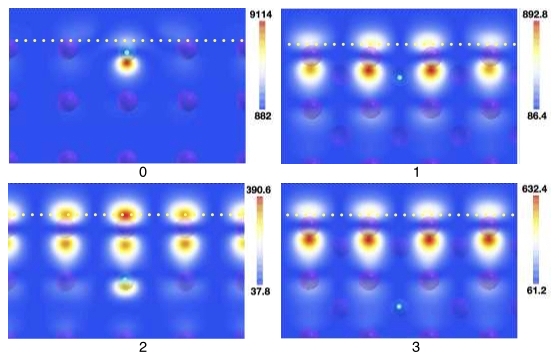}
    \caption{Electronic density of Si donor near the AlAs(110) surface at different depths. The cross-sectional cut is through a (001) plane containing the Al atoms on the surface. When the defect is on the 0th and 2nd layer below the surface, the cutting plane also includes the impurity. Cyan point shows the projected location of the Si impurity (always in the center of the figures). The dotted white line is a guide to the eye indicating the surface (approximately at the positions of the surface As atoms). The units are electrons per \AA$^3 \times 10^{-6}$. }
    \label{ED_001_Si_Slab}
\end{figure*}

\begin{figure*}
    \centering
    \includegraphics[width=\textwidth]{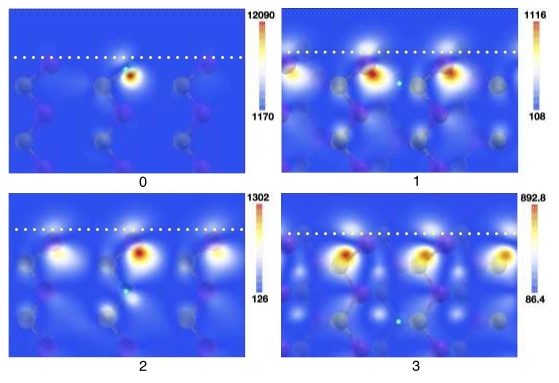}
    \caption{Electronic density of Si donor near the AlAs(110) surface at different depths. The cross-sectional cuts are ($\overline{1}10$) planes containing the surface Al atoms. When the impurity is located at the 0th and 2nd planes, these planes also include the impurity, otherwise they are the nearest ($\overline{1}10$) planes to the impurity. Cyan point shows the projected location of the Si impurity (always in the center of the figure). The dotted white line is a guide to the eye indicating the surface (approximately at the positions of the surface As atoms).  The units are electrons per \AA$^3 \times 10^{-6}$.
 }
 \label{ED_-110_Si_Slab}
\end{figure*}

\begin{figure*}
    \centering
    \includegraphics[width=\textwidth]{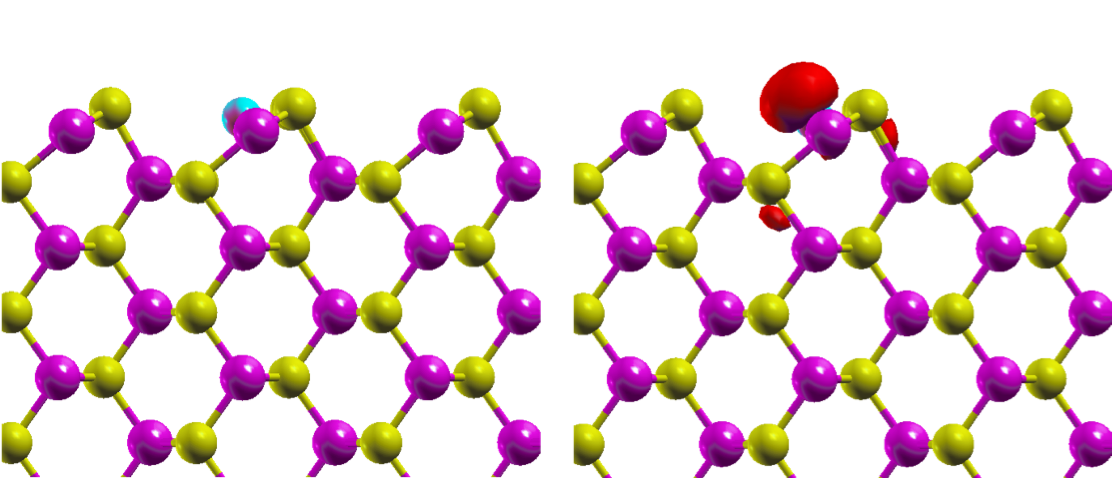}
    \caption{Optimized geometry (left) and 3D plot of a LDOS isosurface (right) of a negatively charged Si atom in the AlAs (110) surface. The LDOS is integrated over an energy range that includes only the impurity state and the isosurface corresponds to 10\% of the maximum value. Here we show a side view corresponding to a ($\bar{1}10$) plane and we can see that the LDOS is highly localized at the dangling bond of the Si atom. The isosurface is shown in red and the color code for the atoms is the same as of Fig. \ref{fig:slab_geometry}.}
\end{figure*}

\begin{figure*}
    \centering
    \includegraphics[width=0.75\textwidth]{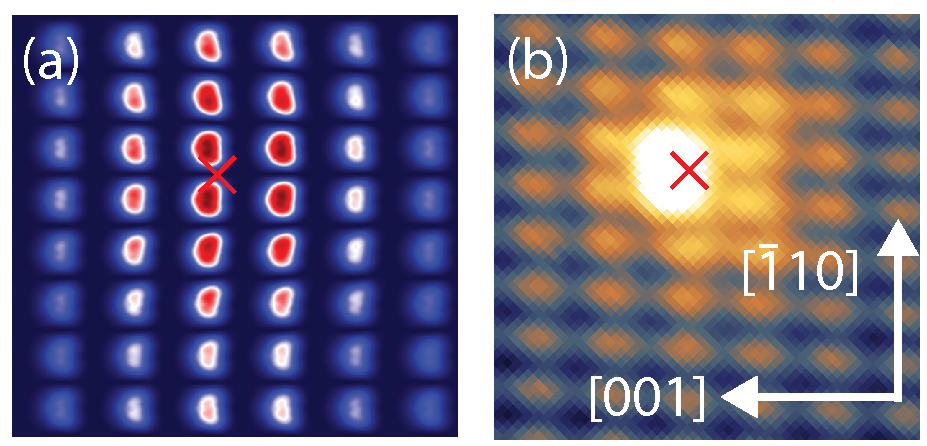}
    \caption{Simulated STM image from DFT calculations (a) and an 2.5 $\times$ 2.5 nm$^2$ empty state STM images (b) of a Si donor three layers below the surface. The red cross in each image marks the location of the Si donor.}
\end{figure*}